\documentclass[sigconf,screen]{acmart}

\usepackage[utf8]{inputenc}
\usepackage[T1]{fontenc}

\usepackage{multirow}
\usepackage{multicol}
\usepackage{adjustbox}
\usepackage{gensymb}

\usepackage{tabularx}
\newcolumntype{C}[1]{>{\centering\arraybackslash}p{#1}}

\usepackage{booktabs} 
\usepackage{mathtools} 
\usepackage{graphicx}
\usepackage{epsfig}

\usepackage{bbding}
\usepackage{pifont}
\usepackage{wasysym}
\usepackage{amssymb}
\usepackage{arydshln}
\usepackage{balance}
\usepackage{subfig}

\usepackage{bookmark}

\begin{document}

\copyrightyear{2019} 
\acmYear{2019} 
\setcopyright{acmlicensed}
\acmConference[ICMR '19]{International Conference on Multimedia Retrieval}{June 10--13, 2019}{Ottawa, ON, Canada}
\acmBooktitle{International Conference on Multimedia Retrieval (ICMR '19), June 10--13, 2019, Ottawa, ON, Canada}
\acmPrice{15.00}
\acmDOI{10.1145/3323873.3325047}
\acmISBN{978-1-4503-6765-3/19/06}

\settopmatter{printacmref=true}
\fancyhead{}

\graphicspath{{images/}}

\title{A Benchmark of Visual Storytelling in Social Media~\textsuperscript{*}}

\thanks{* Please cite the ACM ICMR 2019 version of this paper.}

    \author{Gonçalo Marcelino}
    \affiliation{%
      \institution{NOVALINCS}
      \streetaddress{}
      \city{Uni. NOVA de Lisboa, Portugal} 
    }
    \email{goncalo.bfm@gmail.com}

    \author{David Semedo}
    \affiliation{%
      \institution{NOVALINCS}
      \streetaddress{}
      \city{Uni. NOVA de Lisboa, Portugal} 
    }
    \email{df.semedo@campus.fct.unl.pt}
    
    \author{Andr\'e Mour\~ao}
    \affiliation{%
      \institution{NOVALINCS}
      \streetaddress{}
      \city{Uni. NOVA de Lisboa, Portugal} }
    \email{a.mourao@campus.fct.unl.pt}

    \author{Saverio Blasi}
    \affiliation{
      \institution{BBC Research and Development}
      \city{London, UK} }
    \email{saverio.blasi@bbc.co.uk}

    \author{Marta Mrak}
    \affiliation{%
      \institution{BBC Research and Development}
      \city{London, UK} }
    \email{marta.mrak@bbc.co.uk}
    
    \author{João Magalhães}
    \affiliation{%
      \institution{NOVALINCS}
      \city{Uni. NOVA de Lisboa, Portugal} }
    \email{jm.magalhaes@fct.unl.pt}

\renewcommand{\shortauthors}{G. Marcelino et al.}

\begin{abstract}
Media editors in the newsroom are constantly pressed to provide a \textit{"like-being there"} coverage of live events.
Social media provides a disorganised collection of images and videos that media professionals need to grasp before publishing their latest news updated. 
Automated news visual storyline editing with social media content can be very challenging, as it not only entails the task of finding the right content but also making sure that news content evolves coherently over time. 
To tackle these issues, this paper proposes a benchmark for assessing social media visual storylines. 
The SocialStories benchmark, comprised by total of 40 curated stories covering sports and cultural events, provides the experimental setup and introduces novel quantitative metrics to perform a rigorous evaluation of visual storytelling with social media data.
\end{abstract}


\maketitle

\section{Introduction}
Editorial coverage of events is often a challenging task, in that media professionals need to identify interesting stories, summarise each story, and illustrate the story episodes, in order to inform the public about how an \textit{event} unfolded over time. Thanks to its widespread adoption, social media services offer a vast amount of available content, both textual and visual, and is therefore ideal to support the creation and illustration of these event stories~\cite{paulussen2008user,Hu:2012:BNT:2207676.2208672,Doggett2016IdentifyingEN,Tolmie2017SupportingTU, delgado2010automated}.

\begin{figure}[h]
    \centering
    \includegraphics[width=0.9\linewidth]{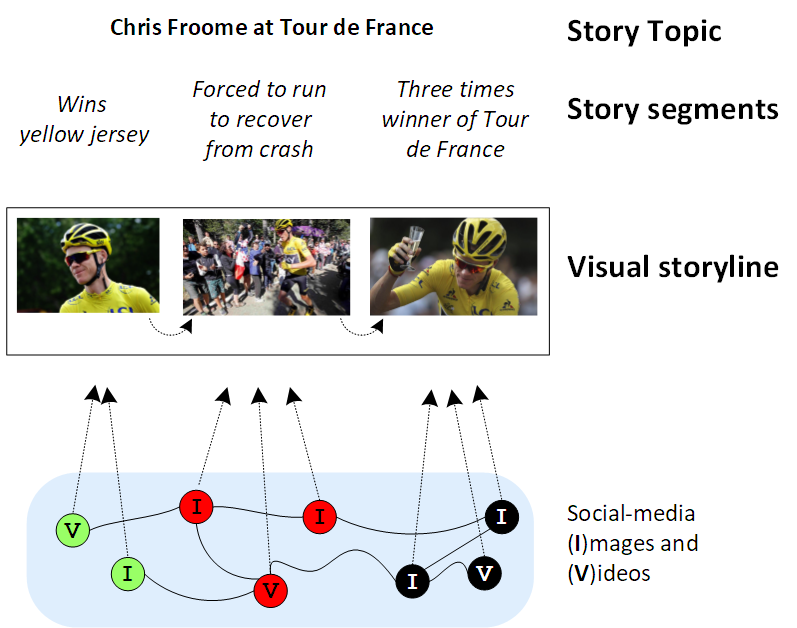}
    \caption{Visual storyline editing task: a news \textit{story topic} and \textit{story segments} can be illustrated by social media content.}
    \label{fig:social-visual-story}
\end{figure}

The timeline of an event, e.g. a music festival, a sport tournament or a natural disaster~\cite{Sakaki:2010:EST:1772690.1772777}, contains visual and textual pieces of information that are strongly correlated. There are several ways of presenting the same event, by covering specific \textit{storylines}, each offering different perspectives. These storylines, illustrated in Figure~\ref{fig:social-visual-story}, refer to a story topic and related subtopics, and are structured into \textit{story segments} that should describe narrow occurrences over the course of the event. More formally, we define a \emph{Visual Storyline} as a sequence of segments, referring to an event topic, with each segment being defined by a textual description and comprising an image or a video.

In Figure~\ref{fig:social-visual-story} we illustrate the newsroom workflow tackled by this paper: once the \textit{story topic} and \textit{story segments} are created, the media editor selects images/videos from social media platforms and organise the retrieved content according to a coherent narrative, i.e. the \textit{visual storyline}.
Many social media platforms, including \href{https://twitter.com/}{Twitter}, \href{https://flickr.com/}{Flickr} or \href{https://youtube.com/}{YouTube} provide a stream of multimodal social media content, naturally yielding an unfiltered event timeline. 
These timelines can be listened to, mined~\cite{Hu:2012:BNT:2207676.2208672, conf/icwsm/ChuaA13, ICWSM112885}, and exploited to gather visual content, specifically image and video~\cite{McParlane2014, Schinas2015}.

\begin{table*}[t]
  \centering
\caption{Dataset statistics for each event, including both terms and hashtags. The event and crawling time spans are shown.}
\label{tab:selected_keywords}
\begin{adjustbox}{center}
    	\resizebox{\textwidth}{!}{%
    \begin{tabular}{ lcclllrl }
    \toprule
    \textbf{Event} & \textbf{Stories} & \textbf{Docs} & \textbf{Docs w/images} & \textbf{Docs w/videos}  & \textbf{Crawling span}  & \multicolumn{2}{l}{\textbf{Crawling seeds}}\\ \hline
    \multirow{2}{*}{EdFest} & 
                         \multirow{2}{*}{20} & \multirow{2}{*}{82,348} &  Twitter: 15439 & Twitter: 3690 & From: 2016-07-01  & Terms & Edinburgh Festival,
                         Edfest,
                         Edinburgh Festival 2016,
                         Edfest 2016\\ 
                         & & & Flickr: 5908 & Youtube: 293 & Until: 2017-01-01  & Hashtags & \#edfest, \#edfringe, \#EdinburghFestival, \#edinburghfest\\ \hline
    \multirow{2}{*}{TDF} & \multirow{2}{*}{20} & \multirow{2}{*}{325,074} & Twitter: 34865 & Twitter: 8677 & From: 2016-06-01  & Terms & le tour de france,
                         le tour de france 2016,
                         tour de france\\ 
                         & & & Flickr: 6442 & Youtube: 983 & Until: 2017-01-01 & Hashtags & \#TDF2016, \#TDF\\
    \bottomrule
    \end{tabular}}%
\end{adjustbox}
\end{table*}

The primary contribution of this paper is the introduction of a quality metric to assess visual storylines. 
This metric is designed to evaluate the quality of an automatically illustrated storyline, based on computational aspects that attempt to mimic the human-driven editorial perspective. 
The quality metric focuses on the \textit{relevance of each individual segment's illustration} and on the general flow of the storyline, i.e. the transitions from segment's illustration to segment's illustration. The second contribution of this paper is a social media dataset with news storylines to allow the research of visual storytelling with social media data. The benchmark, developed for TRECVID2018~\cite{2018trecvidawad}, provides a rigorous setting to research the underpinnings of visual social-storytelling. The most relevant aspect of this benchmark is the realistic nature of the storylines, that mimic the newsroom media editorial process: some stories were manually investigated and inferred from social media and other stories were constructed from existing news articles.

\section{Social Stories Benchmark\protect\footnote{\url{https://novasearch.org/datasets/}.}}
Assessing the success of news visual storyline creation is a complex task. In this section, we address this task and propose the SocialStories benchmark.
Visual storytelling datasets like~\cite{huang2016visual} and~\cite{KimMS15} contain sequences of image-caption pairs, that capture a specific activity, e.g., "\textit{playing frisbee with a dog}".
A characteristic of these stories is that the sequence of visual elements is very coherent (visually and textually), which is highly unlikely to occur in social media.
Hence, these stories do not match the ones a journalist or a media professional needs to create and illustrate on a daily basis.
This highlights the importance of a suitable experimental test-bed for the task at hand.

The SocialStories benchmark provides the experimental setup and metrics to perform a rigorous evaluation of the task of creating visual storylines from social media data.
The core aspects of the SocialStories benchmark are:
\begin{itemize}
    \item \textbf{Storylines:} We created three types of storylines: news article, investigative topics, and review topics. These were either obtained from newswire articles, or created manually through data inspection.
    
    \item \textbf{Assessing Visual Storylines Quality:} Assessing the quality of a sequence of information is a novel and challenging task. We propose a new metric that assesses the quality of a visual storyline in terms of its relevance and transition between segment illustrations.

\end{itemize}
The following sections detail the types of storylines considered in the benchmark and proposes a story quality assessment metric.

\subsection{SocialStories: Event Data and Storylines}
To enable social media visual storyline illustration, a data collection strategy was designed to create a suitable corpora, limiting the number of retrieved documents to those posted during the span of the event. 
Events adequate for storytelling were selected, namely those with strong social-dynamics in terms of temporal variations with respect to their semantics (textual vocabulary and visual content). 
In other words, the unfolding of the event stories is encoded in each collection. Events that span over multiple days like music festivals, sports competitions, etc., are examples of good storyline candidates. 
Taking the aforementioned aspects into account, the data for the following events was crawled (Table~\ref{tab:selected_keywords}):
\begin{description}
	\item[The Edinburgh Festival (EdFest)] consists of a celebration of the performing arts, gathering dance, opera, music and theatre performers from all over the world. The event takes place in Edinburgh, Scotland and has a duration of 3 weeks in August.
	\item[Le Tour de France (TDF)] is one of the main road cycling race competitions. The event takes place in France (16 days), Spain (1 day), Andorra (3 days) and Switzerland (3 days).
\end{description}

\subsubsection{Crawling Strategy} Our \textit{keyword-based} approach, consists of querying the social media APIs with a set of keyword terms. Thus, a curated list of keywords was manually selected for each event. Furthermore, hashtags in social media play the essential role of grouping similar content (e.g. content belonging to the same event)~\cite{Laniado:2010:MST:1940281.1940312}. Therefore a set of relevant hashtags grouping content of the same topic was also manually defined. The data collected is detailed in Table~\ref{tab:selected_keywords}. With no loss of generality, Twitter data was used for the experiments reported in the following sections.

\subsubsection{Story Segments} For each event, newsworthy, informative and interesting topics are considered, containing diverse visual material either in terms of low-level visual aspects (colour, backgrounds, shapes, etc.) and/or semantic visual aspects. Each storyline contains 3 to 4 story segments.

\subsection{Visual Storyline Quality Metric}
\label{sec:metric}
Media editors are constantly judging the quality of news material to decide if it deserves being published. The task is highly skillful and deriving a methodology from such process is not straightforward. 
The task of identifying visual material suitable to describe each story segment is, from the perspective of media professionals, highly subjective.
The motivation for why some content may be used to illustrate specific segments can derive from a variety of factors. 
While subjective preference obviously plays a part in this process (which cannot be replicated by an automated process), other factors are also important which come from common practice and general guidelines, and which can be mimicked by objective quality assessment metrics. 

\begin{figure}[t]
    \centering
    \includegraphics[width=0.8\linewidth]{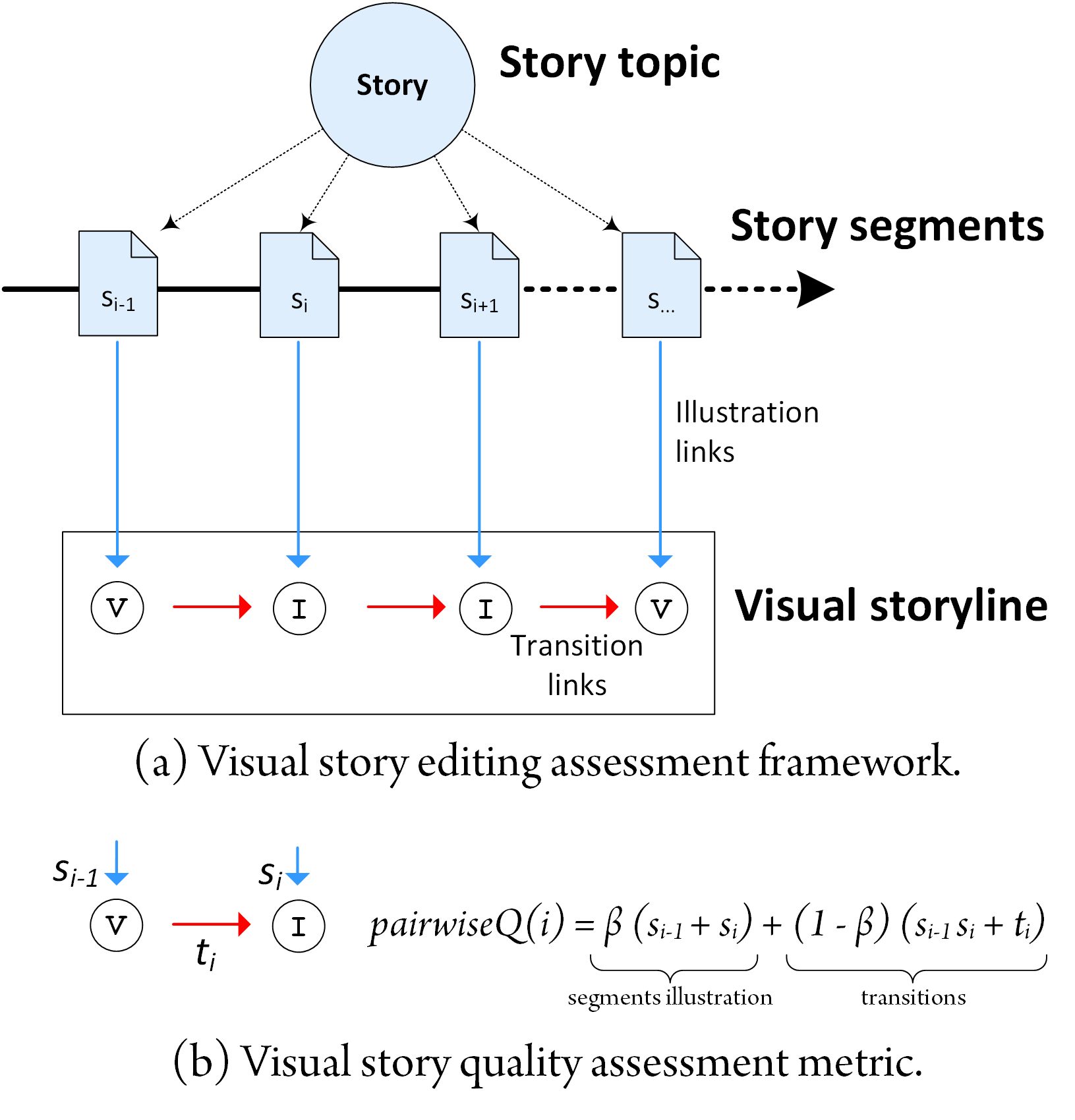}
    \caption{Benchmarking visual storytelling creation.}
    \label{fig:evaluation}
    \vspace{-3mm}
\end{figure}
The first step towards the quantification of visual storyline quality concerns the human-judgement of these different dimensions. 
This is achieved in a sound manner by judging specific objective characteristics of the story -- Figure~\ref{fig:evaluation} illustrates the visual storyline quality assessment framework. 
In particular, storyline illustrations are assessed in terms of \textit{relevance of illustrations} (blue links in Figure~\ref{fig:evaluation}) and \textit{coherence of transitions} (red links in Figure~\ref{fig:evaluation}). 
Once a visual storyline is generated, annotators will judge the relevance of the \textit{story segment illustration} as:
\begin{itemize}
    \item $s_i$=0: the image/video is not relevant to the story segment;
	\item $s_i$=1: the image/video is relevant to the story segment;
	\item $s_i$=2: the image/video is highly relevant to the story segment.
\end{itemize}
Similarly with respect to the \textit{coherence} of a visual storyline, each \textit{story transition} is judged by annotators as the degree of affinity between pairs of story segment illustrations:
\begin{itemize}
    \item $t_i$=0: there is no relation between the segment illustrations;
	\item $t_i$=1: there is a relation between the two segments;
	\item $t_i$=2: there is an appealing semantic and visual coherence between the two segment illustrations.
\end{itemize}
These two dimensions can be used to obtain an overall expression of the \textit{"quality"} of a given illustration for a story of $N$ segments. This is formalised by the expression:
\begin{equation}
Quality=\alpha \cdot s_1 +  \frac{(1-\alpha)}{2(N-1)} \sum_{i=2}^N{pairwiseQ(i)}
\end{equation}
The function $pairwiseQ(i)$ defines quantitatively the perceived quality of two neighbouring segment illustrations based on their relevance and transition:
\begin{align}
pairwiseQ(i)  &= \underbrace{\beta \cdot (s_{i}+ s_{i-1})}_\text{segments illustration} + \underbrace{(1-\beta) \cdot (s_{i-1} \cdot s_{i} + t_{i-1})}_\text{transition}
\end{align}
where $\alpha$ weights the importance of the first segment, and $\beta$ weights the trade-off between \textit{relevance of segment illustrations} and \textit{coherence of transitions} towards the overall quality of the story. 

Given the underlying subjectivity of the task, the values of $\alpha$ or $\beta$ that optimally represents the human perception of visual stories, are in fact average values. Nevertheless, we posit the following two reasonable criteria: (i) illustrating with non-relevant elements ($s_i=0$) completely breaks the story perception and should be penalised. Thus, we consider values of $\beta > 0.5$; and (ii) the first image/video perceived is assumed to be more important, as it should grab the attention towards consuming the rest of the story. Thus, $\alpha$ is a boost to the first story segment $s_1$.
It was empirically found that $\alpha = 0.1$ and $\beta = 0.6$ adequately represent human perception of visual stories editing.

\section{Evaluation}

\subsection{Protocol and Ground-truth} 
\textbf{Protocol.} The goal of this experiment is demonstrate the robustness of the proposed benchmark. Target storylines and segments were obtained using several methods, resulting in a total of 40 generated storylines (20 for each event), each comprising 3 to 4 segments. Ground truth for both relevant segment illustrations, transitions and global story quality were obtained as described in the following section.

\vspace{2mm}
\noindent
\textbf{Ground-truth.} Three annotators were presented with each story title, and asked to rate (i) each segment illustration as relevant or non-relevant, (ii) the transitions between each of the segments, and finally, (iii) the overall story quality. Stories were visualised and assessed in a specifically designed prototype interface. It presents media in a sequential manner to create the right \textit{story} mindset to the user. Using the subjective assessment of the annotators, the score proposed in Section~\ref{sec:metric} was calculated for each story.

\subsection{Quality Metric vs Human Judgement}
\begin{figure*}[t!]
    \centering
    \begin{minipage}{0.2\textwidth}
    \centering
    \includegraphics[width=1.0\linewidth]{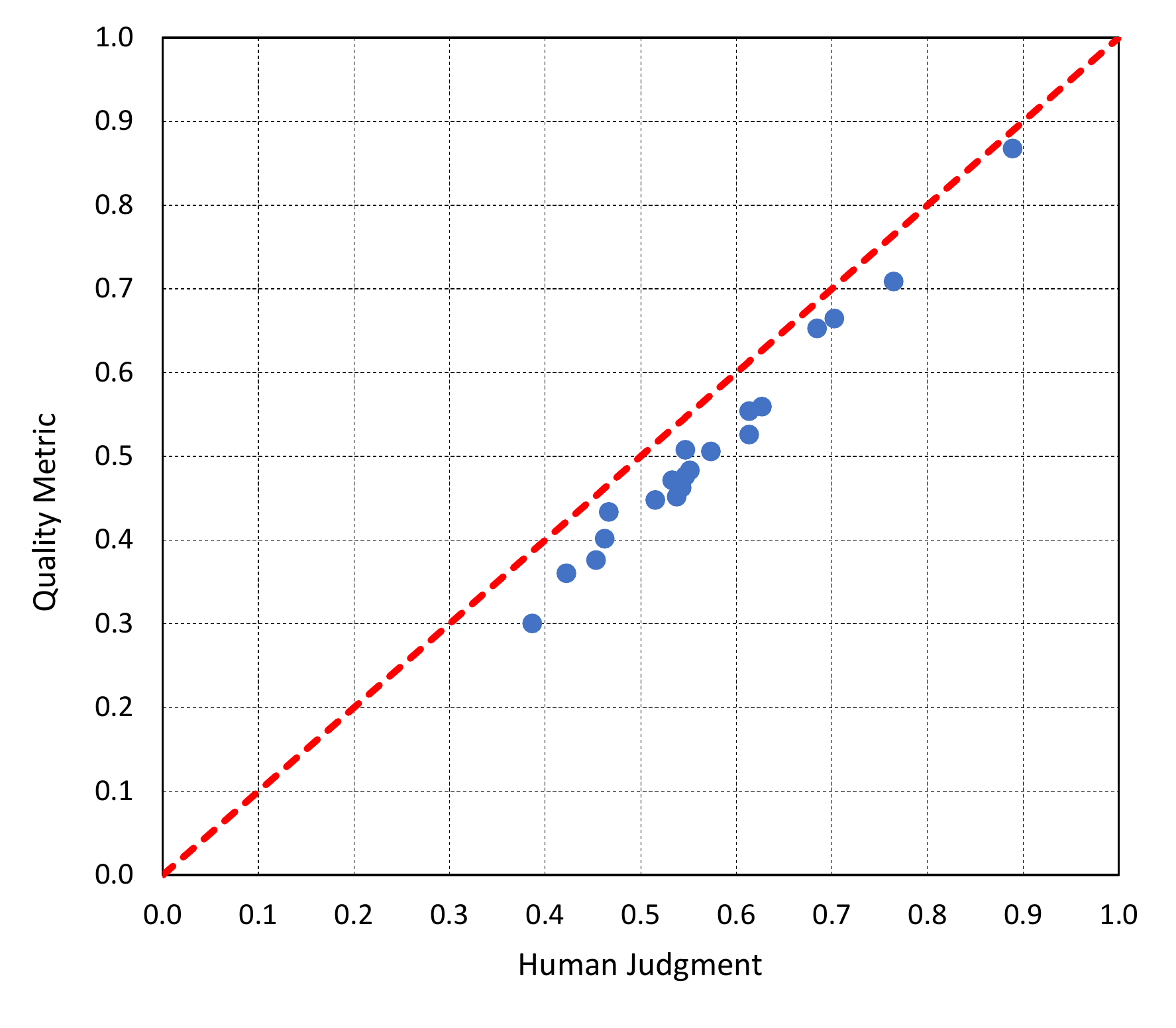}
    \caption{Correlation between human judgement and quality metric. }
    \label{fig:correlation}
    \end{minipage}
    \qquad
    \begin{minipage}{0.75\textwidth}
    \centering
    \subfloat[Illustrations quality]{{\includegraphics[width=0.46\linewidth]{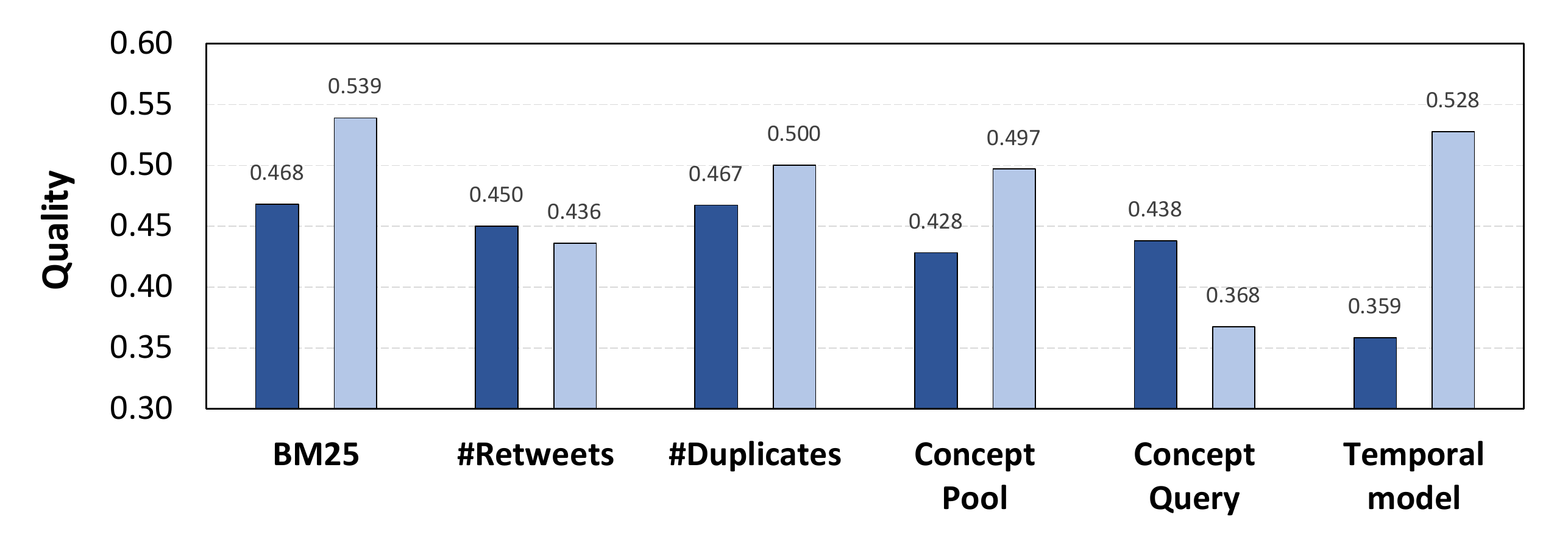} }}%
    \qquad
    \subfloat[Transitions quality.]{{\includegraphics[trim={0 0 0 0},clip,width=0.46\linewidth]{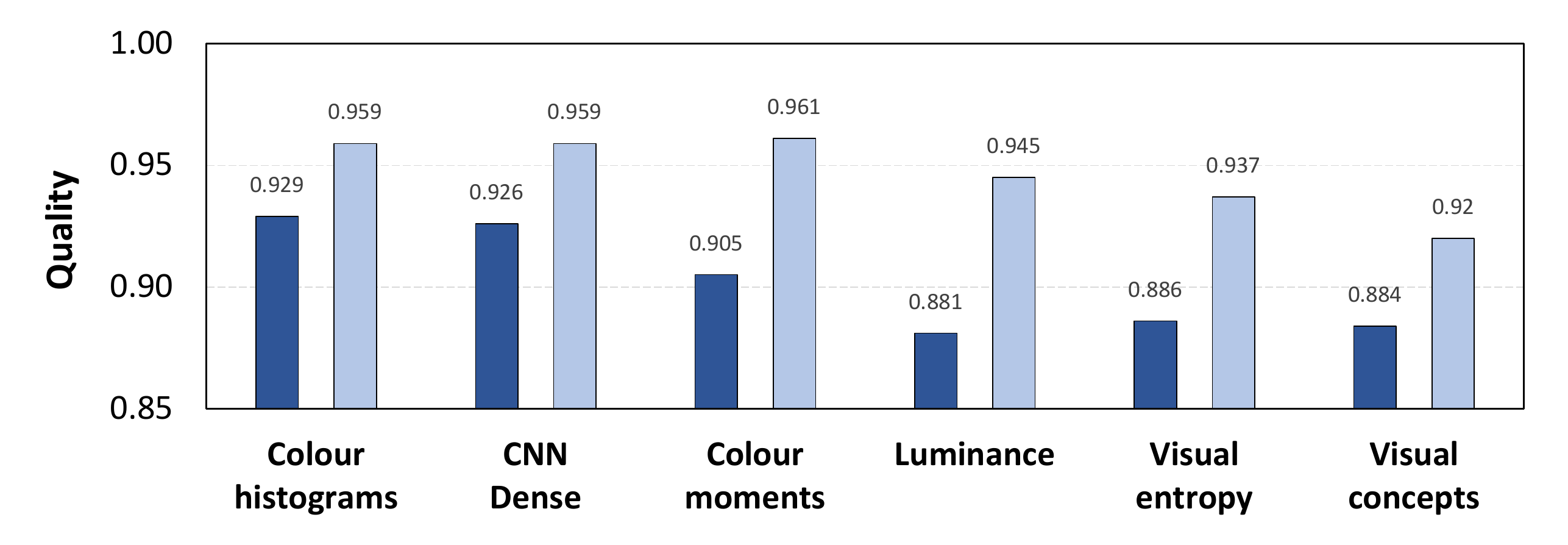}
     }}%
    \caption{Analysis of quality metric in terms of illustration relevance and transition consistency. }%
    \label{fig:illustrations_and_transitions}%
    \end{minipage}
    \vspace{-2mm}
\end{figure*}

In order to test the stability of the metric proposed to emulate the human's perception of visual stories quality, we resorted to crowd-sourcing. To do so, we computed the metric based on the relevance of segments and transitions between segments, and related it to the overall story rating assigned by annotators. Figure~\ref{fig:correlation} compares the annotator rating to the quality metric.
These values show that linear increments in the ratings provided by the annotators were matched by the metric. As can be seen, the relation is strong and relatively stable, which is a good indicator of the metric stability.
Thus, these results show that the metric \emph{Quality} effectively emulates the human perception of visual storyline quality.

\subsection{Automatic Visual Storytelling}

\subsubsection{Results: Illustrations}
Figure~\ref{fig:illustrations_and_transitions} \textbf{(a)} presents the influence of illustrations in the story \emph{Quality} metric introduced in Section~\ref{sec:metric}.

A Text Retrieval baseline (\textit{BM25}) was the best performing baseline for both events. This shows the importance of considering the text of social media documents when choosing the images to illustrate the segments. Approaches based on social signals - (\textit{\#Retweets} and \textit{\#Duplicates}) - also attained good results. Particularly, when inspecting the storylines that result from using the \textit{\#Duplicates} baseline (Figure~\ref{fig:illustrations_and_transitions}), an increase in aesthetic quality of the images selected for illustration can be noticed. This shows that social signals are a powerful indicator of the quality of social media content. However, in scenarios where relevant content is scarce, the approach is hindered by noise. This is especially problematic in cases where there are no strong social signals associated with the available content.

A visual concept detector baseline, based on a pre-trained VGG-16~\cite{SimonyanZ14a} CNN,  did not perform as expected. A \textit{Concept Pool} method selects the image with the 10 most popular visual concepts. It failed often in correctly attributing concepts to the visual content of both datasets. As an example, in TDF, for segments featuring cyclists, unrelated concepts such as  \textit{"unicycle"}, \textit{"bathing\_cap"} or \textit{"ballplayer"} appeared very frequently. However, even though the extracted concepts lack precision, the concepts are extracted consistently among different pieces of content. This improves the performance of the method, since concepts are used to assess similarities in the content. 
The \textit{Concept Query} is a pseudo-relevance feedback-based method that performed even worse.
Hence, the performance of these baselines was lower than that of Text Retrieval. Nevertheless, they were able to effectively retrieve relevant content in situations where relevant content that could be found through text retrieval was scarce or non-existing. 

Finally, we tested a temporal smoothing method \cite{Martins2016}, \textit{Temp. Modeling} to analyse the importance of temporal evidence. It performed the worst on the EdFest test set, while performing  second best for the TDF test set. This high variation in performance is justified by the strong impact the type of story being illustrated has on the behaviour of this baseline. For story segments that take place over the course of the whole event, the approach is not well suited as the probabilities attributed to each segment illustration candidate are virtually identical, and therefore the baseline does not differentiate well between visual elements. However, in story segments with large variations on the amount of content posted per day, the approach provides a noticeably good performance.

Overall, as shown by the results presented in Figure~\ref{fig:illustrations_and_transitions}, generating storylines for Tour de France stories is an easier task than doing so for the Edinburgh Festival stories. 
As a result 4 of the 6 baselines tested performed better in the task of illustrating Tour de France stories.
This happens due to the heterogeneous nature of the media and stories associated with Edinburgh Festival (where very different activities happen) which makes the task of retrieving relevant content more difficult. 
Additionally, the availability of fewer images and videos for Edinburgh Festival accentuates this problem, as there is clearly less data to exploit when creating the storylines.

\subsubsection{Results: Transitions}
Figure~\ref{fig:illustrations_and_transitions}(b) shows the performance of the proposed transitions baselines on the task of illustrating the EdFest and TDF storylines, using the story quality metric introduced in Section~\ref{sec:metric}, calculated based on the judgements of the annotators.
All baselines use the same pool of manually selected relevant visual content, for each segment. 

The \textit{CNN Dense} baseline, minimises distance between representations extracted from the penultimate layer of the visual concept detector. This baseline provided the best performance, highlighting the importance of taking into account semantics when optimising the  quality of transitions. Semantics may not be enough to evaluate the quality of transitions though. In fact, using single concepts as is the case for the \textit{Visual Concepts} baseline provides very poor results, stressing the importance of taking into account multiple aspects.

The second and third best performing baseline focus on minimising the colour difference between images in a storyline: \textit{Colour Histograms} and \textit{Colour Moments}. This supports the assumption that illustrating storylines using content with similar colour palettes is a solid way to optimise the quality of visual storylines.
Conversely, illustrating storylines by selecting images with similar degrees of entropy and luminance, using the \textit{Visual Entropy} and \textit{Luminance} baselines, provided worst results.
Additionally, and similarly to what was observed while assessing the segment illustration baselines, Figure~\ref{fig:illustrations_and_transitions} shows that creating storylines with good transitions is easier for the TDF dataset than for the EdFest dataset.

\section{Conclusions}
This paper addressed the problem of automatic visual story editing using social media data that run in TRECVID2018. Media professionals are asked to cover large events and are required to manually process large amounts of social media data to create event plots and select appropriate pieces of content for each segment. We tackle the novel task of automating this process, enabling media professionals to take full advantage of social media content.
The main contribution of this paper is \textit{a benchmark to assess the overall quality of a visual story} based on the relevance of individual illustrations and transitions between consecutive segment illustrations. 
It was shown that the proposed  experimental test-bed proved to be effective in the assessment of story editing and composition with social media material. 

\vspace{3mm}
\noindent
\textbf{Acknowledgements. } This work has been partially funded by the GoLocal CMU-Portugal project Ref. CMUP-ERI/TIC/0046/2014, by the COGNITUS H2020 ICT project No 687605 and by the project NOVA LINCS Ref. UID/CEC/04516/2013.

\clearpage
\bibliographystyle{ACM-Reference-Format}
\balance
\bibliography{egbib}

\end{document}